\documentclass[12pt]{article}
\usepackage[dvips]{epsfig}
\def\lb{\label}
\def\be{\begin{equation}}
\def\ee{\end{equation}}
\def\ba{\begin{eqnarray}}
\def\ea{\end{eqnarray}}

\def\bb{\bibitem}

\def\e{{\rm e}}
\def\M{{\cal M}}
\def\A{{\cal A}}
\def\sA{\mbox{\LARGE $a$}}
\def\E3{($\alpha^2=3$) EMD}
\def\tb{\bar{\theta}}
\def\bgb{\bar\beta}
\def\bgd{\bar\delta}

\renewcommand{\theequation}{\arabic{section}.\arabic{equation}}
\begin{document}
\begin{titlepage}
\date{}
\title{
   \begin{flushright} \begin{small}
     LAPTH-1042/04  \\
  \end{small} \end{flushright}
\vspace{.5cm}
{\bf Non-asymptotically flat, non-AdS dilaton black holes} }
\author{
G\'erard Cl\'ement\thanks{ Email: gclement@lapp.in2p3.fr} and
C\'edric Leygnac\thanks{ Email: leygnac@lapp.in2p3.fr} \\ \\ {\small
Laboratoire de  Physique Th\'eorique LAPTH (CNRS),} \\ {\small
B.P.110, F-74941 Annecy-le-Vieux cedex, France}}

\maketitle
\begin{abstract}

We show that previously known non-asymptotically flat static black
hole solutions of Einstein-Maxwell-dilaton theory may be obtained as
near-horizon limits of asymptotically flat black holes. Specializing
to the case of the dilaton coupling constant $\alpha^2 = 3$, we
generate from the non-asymptotically flat magnetostatic or
electrostatic black holes two classes of rotating dyonic
black hole solutions. The rotating dyonic black holes of the 
``magnetic'' class are dimensional reductions of the five-dimensional 
Myers-Perry black holes relative to one of the azimuthal angles, while 
those of the ``electric'' class are twisted dimensional reductions of
rotating dyonic Rasheed black strings. We compute the quasi-local mass
and angular momentum of our rotating dyonic black holes, and show
that they satisfy the first law of black hole thermodynamics, as well
as a generalized Smarr formula. We also discuss the construction of
non-asymptotically flat multi-extreme black hole configurations.
   
\end{abstract}
\end{titlepage}
\setcounter{page}{2}
\section{Introduction}

The investigation of black hole properties in general relativity has
long been restricted to the case of asymptotically flat or
asymptotically (anti-)de Sitter ((A)dS) black holes. To our knowledge,
non-asymptotically flat, non-AdS black holes in four (and higher)
dimensions were first found in \cite{CHM} as spherically symmetric
solutions to Einstein-Maxwell-dilaton theory. The properties of such
non-asymptotically flat solutions to Einstein-Maxwell-dilaton-axion
theory in four dimensions were extensively studied in \cite{newdil1}.
Similar non-asymptotically flat topological black holes in four
dimensions were found in \cite{CaJiSo98}, and extended to higher
dimensions in \cite{CaZh01}. The aim of the present work is, first to
reinvestigate four-dimensional non-asymptotically flat, spherically
symmetric dilaton black holes for general dilatonic coupling,
second to analyze in more detail the special case with dilaton
coupling constant $\alpha^2 = 3$.

A motivation to investigate non-asymptotically flat, non-AdS  black
holes is that these might lead to possible extensions of AdS/CFT
correspondence. Indeed, it has been speculated that the linear dilaton
spacetimes, which arise as near-horizon limits of dilatonic black
holes, might exhibit holography \cite{ABKS}. Another  motivation is
that such solutions may be used to extend the range of validity of
methods and tools originally developed for, and tested in the case of,
asymptotically flat or asymptotically AdS black holes. Specifically,
we shall show that the quasi-local energy approach
\cite{BY,HaHo95,Chen,Booth} may be applied successfully to the
computation of the mass and angular momentum of non-asymptotically
flat rotating black holes. We shall also show that such black holes
follow the first law of black-hole thermodynamics, originally
formulated in the case of asymptotically flat black holes \cite{BCH}.

In the next section we recover the non-asymptotically flat (NAF)
static black hole solutions of Einstein-Maxwell-dilaton (EMD) theory as
near-horizon, near-extreme limits of asymptotically flat black holes,
and we discuss briefly their properties. Section 3 is devoted to
multicenter solutions of EMD; these are found to fall in two classes,
one of which includes both asymptotically flat and NAF multi-extreme
black holes. The computation of the quasilocal mass of the NAF static
black holes is recalled in Sect. 4.

The case with dilaton coupling constant $\alpha^2 = 1$ has been
considered in more detail in \cite{newdil1}, where the NAF static
black holes have been extended to rotating black hole solutions of 
Einstein-Maxwell-dilaton-axion gravity (EMDA). Here we shall consider the 
case $\alpha^2 = 3$, which is a dimensional reduction of five-dimensional
Kaluza-Klein theory. Unlike the case of four-dimensional general
relativity, the five-dimensional vacuum Einstein equations admit
solutions with event horizons of various topologies \cite {CaGa01}:
the static Gibbons-Wiltshire black string (topology $S^2 \times R$) 
family \cite{GW}, extended to rotating black strings by Rasheed
\cite{Rasheed}; the static Tangherlini black hole (topology $S^3$) 
\cite{Tang}, extended to rotating black holes by Myers and Perry
\cite{MP5}; and the Emparan-Reall rotating black rings (topology $S^2
\times S^1$) \cite{ER}. These are all asymptotically flat
in five-dimensions. The NAF \E3 static black holes and their
rotating and dyonic generalizations will turn out to correspond to
special dimensional reductions of these asymptotically flat 
five-dimensional black holes. 

In Sect. 5, we use the group $SL(3,R)$ of invariance transformations
of the stationary sector of \E3 to generate NAF rotating magnetic
black holes. We find that these are a dimensional reduction of a
subclass of Myers-Perry black holes. Conversely, we find in 
Sect. 6 that the dimensional reduction of the
generic Myers-Perry black hole depending on three parameters (mass and
two angular momenta) leads to a wider class of NAF rotating dyonic
black holes. In the extreme case, these may be further generalized to 
regular multicenter configurations. We then compute the quasilocal
mass and angular momentum of our rotating dyonic black holes, and
check the validity of the generalized first law of thermodynamics
\cite{Rasheed}. Finally, we use in Sect. 7 electromagnetic duality 
to generate from this ``magnetic'' family a dual ``electric'' sector
of NAF rotating dyonic black holes, which are found to correspond to a
twisted dimensional reduction of Rasheed rotating dyonic black strings 
with a NUT charge balancing the magnetic charge. We conclude in Sect. 8.      

\setcounter{equation}{0}
\section{Static NAF dilaton black holes}

Consider EMD theory, defined by the action
\be \lb{ac}
S = \frac{1}{16\pi}\int d^4x\sqrt{|g|}\left\{R -
2\partial_\mu\phi\partial^\mu\phi
-\e^{-2\alpha\phi}F_{\mu\nu}F^{\mu\nu}\right\} ,
\ee
where $F=d\A$, and $\alpha$ is the dilaton coupling constant. Special
values of $\alpha$ are $\alpha = 0$, corresponding to
Einstein-Maxwell theory, $\alpha = 1$, which is a truncation of the
bosonic sector of $D=4,\;{\cal N}=4$ supergravity, and $\alpha =
\sqrt{3}$, which corresponds to a dimensional reduction of
five-dimensional Einstein gravity with a spacelike Killing vector
(Kaluza-Klein theory). The static, spherically symmetric,
asymptotically flat black hole solutions of EMD were found in
\cite{GM} and rediscovered in \cite{GHS}. They exist in two versions,
electrostatic or magnetostatic, related to each other by the
electro-magnetic duality transformation
\be\lb{emdual}
(\phi, F) \to (\hat{\phi} = -\phi, \,\hat{F} =
\pm\e^{-2\alpha\phi}\tilde{F}),
\ee
with\footnote{Here
$E^{\mu\nu\lambda\tau} \equiv
|g|^{-1/2}\varepsilon^{\mu\nu\lambda\tau}$, with $\varepsilon^{1234}
= +1$, where $x^4 = t$ is the time coordinate.} ${\tilde
F}^{\mu\nu}=\frac{1}{2}E^{\mu\nu\lambda\tau}F_{\lambda\tau}$. The
electric black hole solutions are
\ba\lb{ghs}
ds^2 & = & -\frac{(r-r_-)^{\gamma}(r-r_+)}{r^{1+\gamma}}\,dt^2 +
\nonumber \\ & & \quad + \,
\frac{r^{1+\gamma}}{(r-r_-)^{\gamma}(r-r_+)}\bigg[\,dr^2 +
(r-r_-)(r-r_+)\,d\Omega^2\bigg], \\ F & = &
\frac{Qe^{2\alpha\phi_{\infty}}}{r^2}\,dr\wedge dt, \qquad
e^{2\alpha(\phi-\phi_{\infty})} =
\bigg(1-\frac{r-}r\bigg)^{1-\gamma}, \nonumber
\ea
with
\be
\gamma = \frac{1-\alpha^2}{1+\alpha^2}.
\ee
The black hole parameters $r_+$ (the location of the event horizon)
and $r_-$ (for $\alpha \neq 0$, the spacelike singularity, $0<r_- <
r_+$) are related to the physical parameters $\M$ (mass) and $Q$
(electric charge) by
\be
\M=\frac{r_++\gamma r_-}{2}, \quad Q=e^{-\alpha\phi_{\infty}}
\sqrt{\frac{1+\gamma}2}\sqrt{r_+ r_-}\,.
\ee

The solutions (\ref{ghs}) have been proved to be the only
electrostatic, asymptotically flat regular black hole solutions for
$\alpha = 1$ \cite{MuA} and, more recently, for arbitrary $\alpha$
\cite{MS}. However this proof leaves open the possibility, for
$\alpha \neq 0$, of NAF electrostatic black
holes, generalizing the $\alpha = 1$ linear dilaton black holes
\cite{GS, newdil1}. These have previously been found by a direct
solution of the field equations of EMD \cite{CHM} (see also
\cite{GuSe95, KiPa97}). Here we shall recover these
non-asymptotically flat black holes by taking the near-extreme,
near-horizon limit of the asymptotically flat solutions (\ref{ghs}).
Define three new black-hole parameters $r_0$, $b$ and $\nu$ by
\be
\quad r_- = \epsilon^{-\alpha^2}r_0,  \quad
r_+=\epsilon^{-\alpha^2}r_0+ \epsilon b, \quad \phi_\infty =
\alpha^{-1}\ln\nu -\alpha\ln \epsilon,
\ee
where the dimensionless parameter $\epsilon$ shall eventually be
taken to zero, and transform the $(t,r)$ coordinates to
\be
t = \epsilon^{-1}\bar{t}, \quad r= \epsilon^{-\alpha^2}r_0 +
\epsilon\bar{r}.
\ee
Finally take the limit $\epsilon\rightarrow0$ and relabel $\bar{t} \to t$,
$\bar{r} \to r$, which yields
\ba
ds^2 & = & -\frac{r^{\gamma}(r-b)}{r_0^{1+\gamma}}\,dt^2 +
\frac{r_0^{1+\gamma}}{r^{\gamma}(r-b)}\bigg[dr^2 +
r(r-b)d\Omega^2\bigg], \lb{nbhg}\\ F & = &
\sqrt{\frac{1+\gamma}2}\,\frac{\nu}{r_0}\,dr\wedge dt\,, \qquad
e^{2\alpha\phi} = \nu^{2}\bigg(\frac{r}{r_0}\bigg)^{1-\gamma}.
\lb{nbhe}
\ea
The magnetic dual version also exists, with the same metric supported
by the magnetic and dilaton fields
\be
F = \sqrt{\frac{1+\gamma}2} \frac{r_0}{\nu}
\sin\theta\,d\theta\wedge\,d\varphi, \qquad e^{2\alpha\phi} =
\nu^{-2}\bigg(\frac{r}{r_0}\bigg)^{\gamma-1}. \lb{nbhm}
\ee

These NAF solutions of EMD depend on three parameters, $\nu$ which
accounts for the invariance under dilaton rescalings $\phi  \to \phi$
$+$ constant, $r_0 > 0$ which sets the overall scale and is related to
the electric charge $Q$ according to
\be\lb{Q}
Q=\frac1{4\pi}\int\e^{-2\alpha\phi}
F^{0r}\sqrt{|g|}\,d\Omega=\sqrt{\frac{1+\gamma}2}\,\frac{r_0}{\nu},
\ee
and the horizon radius $b$, which we expect to be proportional to the
black hole mass $\M$, the exact computation (recalled in Sect. 4)
giving \cite{CHM}
\be
\M = \frac{(1-\gamma)b}4.
\ee
For $\alpha^2 = 0$ ($\gamma=1$), the family of solutions
(\ref{nbhg})-(\ref{nbhm}) depending on $b$ correspond to different
parametrizations of the Bertotti-Robinson spacetime $AdS_2\times S^2$
generated by a homogeneous electric or magnetic field. Being related
to the $b=0$ solution by global coordinate transformations, these
solutions are not black holes, and accordingly their mass vanishes.
For $\alpha^2\rightarrow\infty$ ($\gamma=-1$), the dilaton and
electromagnetic fields decouple and the solutions (\ref{nbhg}) reduce
to the Schwarzschild black holes with the appropriate mass $\M =
b/2$. Thus, this family of non-asymptotically flat black hole
solutions interpolate continuously between the Bertotti-Robinson and
Schwarzschild solutions.

The Penrose diagrams corresponding to the different values of $b$
($b<0$, $b=0$ and $b>0$) are shown in Fig.\ 1 for $0<\alpha^2<1$
($0<\gamma<1$), Fig.\ 2 for $\alpha^2=1$ ($\gamma=0$) and Fig.\ 3 for
$\alpha^2>1$ ($\gamma<0$). The Ricci scalar
\be
R = -\frac{1-\gamma^2}2\frac{(r-b)r^{\gamma-2}}{r_0^{1+\gamma}}
\ee
vanishes at spatial infinity $r\to\infty$ and on the horizon, and is
singular for $r=0$. For $\alpha^2<1$, spatial infinity is conformally
time-like, as in the $AdS$ case, leading to a Penrose diagram for the
black-hole case $b>0$ which is similar to that of the
three-dimensional BTZ static black hole \cite{BTZ}. For $\alpha^2\ge1$, spatial
infinity is conformally null, with a Schwarzschild-like Penrose
diagram in the black-hole case. The extreme black holes ($b=0$) are
all singular, the singularity being null for $\alpha^2\le1$ and
timelike for $\alpha^2>1$.

\setcounter{equation}{0}
\section{Extreme case: multicenter solutions}

As in the special case $b=0$  of the linear dilaton \cite{newdil1},
the extreme $b=0$ solutions have a conformally flat spatial metric,
which suggests that they can be linearly superposed to yield
multicenter solutions. To show this, let us follow the reduction of
the electrostatic sector of EMD to a three-dimensional
self-gravitating $\sigma$ model carried out in \cite{GaGaKe95}. The
dimensional reduction, achieved by
\be
ds^2 = -f\,dt^2 + f^{-1}\gamma_{ij}\,dx^i\,dx^j, \quad F_{i0} =
\frac1{\sqrt2}\,\partial_i v,
\ee
reduces the original 4-dimensional EMD equations to a
three-dimensional problem deriving from the gravity coupled $\sigma$
model action
\be \lb{siga}
S_3 = \int d^3x\sqrt{\gamma}\left\{R_{\gamma} -
G_{AB}(X)\partial_iX^A\partial_jX^B\gamma^{ij}\right\},
\ee
where $R_{\gamma}$ is the Ricci scalar constructed from the 3-dimensional
metric $\gamma_{ij}$, and $G_{AB}(X)$ is the target space metric
\be\lb{tar}
dS^2 = G_{AB}dX^AdX^B = \frac{df^2}{2f^2} -
\frac1f\,\e^{-2\alpha\phi}\,dv^2 + 2\,d\phi^2.
\ee
In the case where the potentials $f$, $v$ and $\phi$ depend on a
single scalar potential $\sigma$, this potential can always
\cite{NeKr69, exact} be chosen
to be harmonic ($\nabla_{\gamma}^2\sigma= 0$). The point ($f$, $v$, $\phi$)
then follows a geodesic in target space, null geodesics leading to a
Ricci-flat, hence flat, reduced 3-space of metric $\gamma_{ij}$
\cite{spat,spat2,ClGa96}. The geodesics for the metric (\ref{tar})
are obtained by solving the system
\ba
f\ddot{f} - \dot{f}^2 & = & f\e^{-2\alpha\phi}\dot{v}^2, \\
(f^{-1}e^{-2\alpha\phi}\dot{v})\,\dot{} & = & 0, \\ 2f\ddot{\phi} & =
& \alpha\e^{-2\alpha\phi}\dot{v}^2.
\ea
It is straightforward to show that, for the choice $\gamma_{ij} =
\delta_{ij}$, there are two kinds of null geodesics.

Null geodesics of the first, generic kind lead (up to a linear
transformation on $\sigma$) to the two-parameter ($c, k$) family of
singular solutions
\ba\lb{ck}
f & = &
\bigg(\frac{ck^{\alpha^2}\e^{\alpha\sigma}}{\sin\sigma}\bigg)^{1+\gamma},
\\ v & = & c\sqrt{1+\gamma}\cot\sigma, \\ \e^{2\alpha\phi} & = &
\bigg(\frac{c}{k\sin\sigma}\bigg)^{1-\gamma}\e^{-(1+\gamma)\alpha\sigma}.
\ea
For $\alpha=1$, these correspond to the electrostatic sector ($\chi =
u = \kappa = 0$) of the type 2 (nondegenerate) solutions found in
\cite{ClGa96} (for instance, the solution (6.17) of \cite{ClGa96}
corresponds to $c = 1/\sqrt{2}, k = \e^{-3\pi/4}$). For $\alpha =
\sqrt{3}$, the lift to five dimensions according to the Kaluza-Klein
ansatz recalled in Sect. 5  leads to the metric
\be
ds_5^2 = k\e^{\sigma/\sqrt{3}}\bigg(-c\sin\sigma\,dt^2 +
2\cos\sigma\,dt\,dx^5 + c^{-1}\sin\sigma\,(dx^5)^2\bigg) +
k^{-2}\e^{-2\sigma/\sqrt{3}}\,d{\bf x}^2,
\ee
which belongs to the electrostatic sector of the class (a) of
solutions found in \cite{spat} (Eq. (25)). Let us recall that the
corresponding multi-center five-dimensional metric generated by the
linear superposition
\be\lb{multi}
\sigma = \sigma_{\infty} + \Sigma_i(c_i/r_i)
\ee
is geodesically complete if all the $c_i$ are negative. Finally, for
$\alpha=0$, the dilaton field does not decouple, so that the
multicenter configurations (\ref{ck}) lead to the family of singular 
solutions of Einstein-Maxwell-massless scalar
field theory:
\be
f = \frac{c^2}{\sin^2\sigma}, \quad v = c\sqrt{2}\cot\sigma, \quad
\phi = -\sigma\,.
\ee

Null geodesics of the second, special kind lead (again up to a linear
transformation on $\sigma$) to the one-parameter family of solutions
\ba
ds^2 & = & -\sigma^{-1-\gamma}\,dt^2 + \sigma^{1+\gamma}\,d{\bf x}^2,
\lb{nugeog}
\\ \A & = & \nu\sqrt{\frac{1+\gamma}2}\,\sigma^{-1}\,dt\,,
\qquad e^{2\alpha\phi} = \nu^2\sigma^{\gamma-1}, \lb{nugeoe}
\ea
where the harmonic function $\sigma$ may for instance be chosen in
the multicenter form (\ref{multi}). As discussed in \cite{GKLTT}, 
for $\sigma_\infty \neq 0$ these are
BPS (supersymmetric) asymptotically flat solutions generalizing the
extreme($r_+=r_-$) asymptotically flat black holes (\ref{ghs}), while
for $\sigma_\infty = 0$ these multicenter solutions, which
generalize (\ref{nbhg})-(\ref{nbhe}) for $b=0$, are NAF. While the
proof of supersymmetry given in \cite{GKLTT} invokes asymptotic
flatness, we conjecture that (as in the special case $\alpha=1$
\cite{KP,newdil1}), these last NAF multicenter solutions are
supersymmetric for all $\alpha$. By electro-magnetic duality one may
derive from these electrostatic solutions the corresponding
multicenter magnetostatic solutions.

\setcounter{equation}{0}
\section{Quasilocal mass}

The mass of the static NAF self-gravitating configurations
(\ref{nbhg})-(\ref{nbhe}) was previously computed in \cite{CHM}. For
the sake of completeness, and to pave the way for a similar
computation of the mass and spin of rotating NAF solutions in the
next two sections, we outline this computation here. It employs the
quasilocal energy approach developed among others by Brown and York
\cite{BY} and put on a firm canonical basis by Hawking and Horowitz
\cite{HaHo95} (see also \cite{Chen, Booth} and references therein).
Consider a spacetime region $M$ bounded by initial and final
spacelike surfaces $\Sigma_{t_1}$ and $\Sigma_{t_2}$, and a timelike
surface $\Sigma^{r}$ (not necessarily at spatial infinity), which we
assume to be orthogonal to the $\Sigma_{t}$. In the canonical $1 + 3$
ADM decomposition \cite{ADM}, the metric and electromagnetic
potential on $M$ are written as
\be\lb{adm}
ds^2 = -N^2\,dt^2 + h_{ij}(dx^i + N^i\,dt)(dx^j + N^j\,dt), \quad \A =
A_0\,dt + A_i\,dx^i,
\ee
where $h_{ij}$ is the induced metric on $\Sigma_t$. The
three-surfaces $\Sigma_t$ and $\Sigma^{r}$ intersect on a two-surface
$S_t^{r}$, with induced metric $\sigma_{\mu\nu} = h_{\mu\nu} -
n_{\mu}n_{\nu}$, where $n^i$ is the unit normal to $\Sigma^{r}$. In
the present case $S_t^{r}$ is a two-sphere of radius $r$, with
\be
\sigma_{ab}\,dx^a\,dx^b = r_0^{1+\gamma}r^{1-\gamma}\,d\Omega^2,
\quad n^r = N = \sqrt{\frac{r^{\gamma}(r-b)}{r_0^{1+\gamma}}}.
\ee
The action (\ref{ac}) in the region $M$, supplemented by boundary
terms necessary to correctly account for Dirichlet boundary
conditions on $\partial M$, can be rearranged after integration by
parts as
\ba\lb{canac}
S & = & \int dt \left[\int_{\Sigma_t}(p^{ij}\dot{h}_{ij} +
p^i\dot{A_i} + p\dot{\phi} - N{\cal H} - N^i{\cal H}_i - A_0{\cal
H}_A) - \right. \nonumber
\\ & & \left. \quad - \oint_{S_t^{r}}(N\epsilon + 2N^i\pi_{ij}n^j +
A_0\Pi^r)\right],
\ea
where $p^{ij}$, $p^i$ and $p$ are the canonical momenta conjugate to
$h_{ij}$, $A_i$ and $\phi$, and ${\cal H}$, ${\cal H}_i$, and ${\cal
H}_A$ are the Hamiltonian, momentum and Coulomb constraints conjugate
to the non-dynamical variables $N$, $N^i$ and $A_0$. These
constraints vanish on shell, the Hamiltonian then reducing to the
surface term
\be\lb{ham}
H = \oint_{S_t^{r}}(N\epsilon + 2N^i\pi_{ij}n^j + A_0\Pi^r),
\ee
where
\be\lb{eps}
\epsilon = \frac1{8\pi} k\sqrt{|\sigma|},
\ee
$k$ being the trace of the extrinsic curvature of $S_t^{r}$ in
$\Sigma_t$,
\be\lb{k}
k = - \sigma^{\mu\nu}D_{\mu}n_{\nu}
\ee
(with $D_{\mu}$ the covariant derivative on $\Sigma_t$), the reduced
momenta $\pi_{ij} =\\ (\sqrt{|\sigma|}/\sqrt{|h|})p_{ij}$ are related
to the extrinsic curvature of $\Sigma_t$,
\be\lb{Kij}
K_{ij} = - \frac1{2N}(\dot{h}_{ij} - 2D_{(i}N_{j)}),
\ee
by
\be
\pi_{ij} = \frac1{16\pi}\sqrt{|\sigma|}(Kh_{ij}-K_{ij}),
\ee
and
\be
\Pi^r = \frac1{4\pi}N\sqrt{|h|}\e^{-2\alpha\phi}F^{rt}.
\ee

The value of the Hamiltonian (\ref{ham}) generically diverges when
the radius $r$ of the 2-sphere $S_t^{r}$ is taken to infinity. The
quasilocal energy or mass is the difference between the value of this
Hamiltonian and that for a suitable reference background or
``vacuum'' evaluated with the same boundary data for the fields
$|\sigma|$, $N$ and $A_0$. Here, the area $4\pi|\sigma|$ of a sphere
of given radius $r$ depends only on the charge parameter $r_0$, hence
the natural choice for the vacuum in a given charge sector ($r_0$
fixed) is the extreme black hole solution with $b=0$. As the electric
potential and dilaton field also depend only on $r_0$ (or vanish in
the magnetic case), it follows that the electric contribution to the
energy cancels between the black hole and background, leaving only
the gravitostatic contribution given by the first term of (\ref{ham})
minus the corresponding background contribution. The computation of
the extrinsic curvature of $S_t^{r}$ gives
\be
k = -(1-\gamma)\sqrt{\frac{r-b}{r_0^{1+\gamma}r^{2-\gamma}}},
\ee
leading, after substraction, to the value for the mass
\be
\M = \frac{(1-\gamma)b}4.
\ee
quoted in Sect. 2.

Now we check agreement with the first law of black hole
thermodynamics
\be\lb{first}
d\M = T\,dS + V_h\,dQ.
\ee
The Hawking temperature $T$ is given by the surface gravity divided
by $2\pi$,
\be
T = \frac1{2\pi}n^i\partial_iN\bigg|_h = b^{\gamma}/4\pi
r_0^{1+\gamma},
\ee
while the black hole entropy is given by a quarter of the horizon
area
\be
S = A_h/4 = \pi r_0^{1+\gamma}b^{1-\gamma}.
\ee
A straightforward computation gives
\be\lb{firsttry}
d\M - T\,dS = \frac{1+\gamma}4\,b\,\frac{dr_0}{r_0}.
\ee

In the magnetostatic case, the electric charge $Q$ is identically
zero, so that the first law holds provided the scale parameter $r_0$
is held fixed during the variation. This makes sense, because we have
defined the mass of a given black hole as the difference between its
energy and the energy of the extreme black hole with the same value
of $r_0$.

In the electrostatic case, the electric charge being proportional to
$r_0/\nu$, the first law again holds if the metric scale $r_0$ and
dilaton scale $\nu$ are both held fixed. There is however another
intriguing possibility. The electric potential associated with (\ref{nbhe}),
\be\lb{gauge}
V = -A_0 = -\nu\sqrt{\frac{1+\gamma}2}\frac{(r-b)}{r_0} - C,
\ee
is defined only up to an additive (gauge) constant $C$. In the case
of, say, the Reissner-Nordstr\"om solution, the potential goes to a
constant at spatial infinity, and there is a preferred gauge in which
the potential vanishes at infinity and the first law (\ref{first}) is
valid. In the present case, the potential (\ref{gauge}) 
diverges at spatial infinity, so that there is {\em a priori} no 
preferred gauge. It is however possible (just as in the case of
charged black holes in $2+1$ gravity \cite{black}) to choose the gauge 
$C$ such that the first law holds when both the parameters $b$ and
$r_0$ are varied (with $\nu$ held fixed). Interestingly enough, with
this choice 
\be\lb{C}
C = \sqrt{\frac{1+\gamma}2}\frac{b\nu}{2r_0}\,,
\ee
both the static versions of the differential first law (\ref{first})
and of the Smarr formula
\be
\M = 2TS + V_hQ
\ee
are satisfied. We suggest that a more careful thermodynamical
treatment of these NAF charged black holes might explain the success
of this apparently {\em ad hoc} procedure.

\setcounter{equation}{0}
\section{Case $\alpha^2 = 3$: rotating magnetic black holes}

Can these static non-asymptotically flat black holes be extended to
rotating black holes? Such an extension was carried out in
\cite{newdil1} for the special case $\alpha^2 = 1$, by taking
advantage of the fact that EMD can be embedded in EMDA
whose stationary sector can
be reduced to a three-dimensional self-gravitating $Sp(4,R)/U(2)$
$\sigma$ model \cite{Ga95}. The strategy used was, first to find the
$Sp(4,R)$ group transformation $U$ generating the linear dilaton
black holes of \cite{newdil1} from the Schwarzschild solution,
then to apply the same transformation to the corresponding Kerr
family of solutions to generate a family of rotating linear dilaton
black holes.

The symmetries of the target space of pure stationary EMD with
general $\alpha$ were studied in \cite{GaGaKe95}, \cite{JeLi95},
where it was shown
that this target space has a symmetric Riemannian space (or $\sigma$
model) structure only in the two cases $\alpha^2 =0$
(Einstein-Maxwell theory, with the target space $SU(2,1)/S(U(1)\times
U(2))$ and $\alpha^2 =3$ (dimensionally reduced Kaluza-Klein theory
with the target space $SL(3,R)/SO(3)$). In the first case, the static
solutions (\ref{nbhg}) and (\ref{nbhe}) or (\ref{nbhm}) all
correspond to Bertotti-Robinson spacetime viewed by different
uniformly accelerating observers. As shown in \cite{kerr}, the
rotating solutions generated from these by the $\sigma$ model
procedure described above again correspond to the same
Bertotti-Robinson spacetime viewed by uniformly rotating observers.
Conversely, this means that the Kerr solution can be generated from
the Schwarzschild solution by a combination of $SU(2,1)$ group
transformations and uniform frame rotations, which can more generally
be used to generate asymptotically dipole solutions of
Einstein-Maxwell theory from asymptotically monopole solutions
\cite{vlad}. We shall focus here on the other case $\alpha^2 = 3$,
corresponding to $\gamma = -1/2$.

It is well-known that EMD theory with $\alpha^2 = 3$ is a dimensional
reduction of 5-dimensional sourceless Kaluza-Klein theory, i.e.
5-dimensional vacuum Einstein gravity
\be\lb{e5}
S = \frac1{16\pi} \int d^5x \sqrt{|g_5|} R_5 \,,
\ee
together with the assumption of a spacelike Killing vector $\partial/\partial
x^5$. Indeed, making the standard Kaluza-Klein ansatz
\be\lb{4+1}
ds_5^2 = e^{2\phi/\sqrt{3}}\,ds_4^2 + e^{-4\phi/\sqrt{3}}\,(dx^5 +
2A_{\mu}dx^{\mu})^2\,,
\ee
and integrating out the cyclic coordinate $x^5$ reduces the action
(\ref{e5}) to the EMD action (\ref{ac}) with $\alpha = \sqrt{3}$. It
follows that the stationary solutions (i.e. solutions with a timelike
Killing vector $\partial_t$) of $\alpha^2 = 3$ EMD are dimensional
reductions of solutions of 5D vacuum gravity with two Killing
vectors. As shown by Maison \cite{maison}, this two-stationary sector
of Kaluza-Klein theory may be reduced to a three-dimensional
self-gravitating $SL(3,R)/SO(3)$ $\sigma$ model. The metric ansatz
appropriate for this five-to-three dimensional reduction is
\be\lb{3+2}
ds_5^{2} = \lambda_{ab} (dx^{a} + \sA^a_i\,dx^i)
(dx^{b} + \sA^b_j\,dx^j) +
{\tau}^{-1}\gamma_{ij}\,dx^i\,dx^j\,,
\ee
where $i=1,...,3$, $a=4,5$ ($x^4 = t$), $\tau = |{\rm
det}(\lambda)|$, and the various fields depend only on the
coordinates $x^i$. Using the $5$--dimensional Einstein equations, the
magnetic--like vector potentials $\sA^a_i$ may be
dualized to the scalar twist potentials $V_a$ according to
\be\lb{dual}
V_{a,i} \equiv |\gamma|^{-1/2}\tau\lambda_{ab}\gamma_{il}\epsilon^{jkl}
\sA^b_{j,k}\,.
\ee
The remaining Einstein equations may then be written in the
$3$--dimensional self-gravitating $\sigma$--model form
\be
(\chi^{-1}\chi^{,i})_{;i} =  0, \quad
R_{ij} =  \frac{1}{4}{\rm
Tr}(\chi^{-1}\chi_{,i}\chi^{-1}\chi_{,j})\,,
\lb{sig}
\ee
where the $3$--metric is $\gamma_{ij}$, and $\chi$ is the unimodular
symmetric $3\times3$ matrix--valued field
\be\lb{chi}
\chi = \left( \begin{array}{ccc}
            \lambda_{ab} - \tau^{-1}V_aV_b & -\tau^{-1}V_a\\
            -\tau^{-1}V_b & -\tau^{-1} \end{array} \right)\,.
\ee
These equations are clearly invariant under $SL(3,R)$ transformations
\be\lb{transf}
\chi \to U^T\chi U,
\ee
which therefore transform a stationary solution of \E3 into another
solution with the same reduced 3-metric $\gamma_{ij}$.

We first consider the magnetic NAF black hole solution (\ref{nbhg}),
(\ref{nbhm}) with $\alpha = \sqrt{3}$ (where we have chosen without
loss of generality $\nu=1$), which leads according to (\ref{4+1}) to
the five-dimensional metric
\be\lb{5m1}
ds_5^2 = -\frac{r-b}{r}\,dt^2 + \frac{r}{r_0}\bigg(dx^5 -
r_0\cos\theta\,d\varphi\bigg)^2 + \frac{r_0}{r-b}\bigg(dr^2 +
r(r-b)\,d\Omega^2\bigg) \,.
\ee
The resulting representative matrix
\be\lb{chim}
\chi_m = \left( \begin{array}{ccc} -\frac{r-b}{r} & 0 & 0
\\0 & -\frac{br}{r_0(r-b)} & \frac{r}{r-b} \\ 0 & \frac{r}{r-b} &
-\frac{r_0}{r-b}
\end{array}\right)
\ee
is a target space geodesic \cite{spat}
\be\lb{targeo}
\chi = \eta\e^{A\sigma},
\ee
with the harmonic potential
\be\lb{harm}
\sigma=-\frac{r_0}{b}\ln\left|\frac{r-b}{r}\right| \,,
\ee
and the constant matrices $\eta$ and $A$
\be
\eta_m =\left(
\begin{array}{ccc}
-1&0&0\\ 0&-\frac{b}{r_0}&1\\ 0&1&0
\end{array}
\right), \quad A_m=\left(
\begin{array}{ccc}
-\frac{b}{r_0}&0&0\\ 0&\frac{b}{r_0}&-1\\ 0&0&0
\end{array}
\right)\,. \label{mA}
\end{equation}

The metric (\ref{5m1}) is regular on the axis $\sin\theta=0$ provided
$x^5$ is periodic with period $4\pi r_0$ \cite{GP}. Making the coordinate
transformation
\be
r = \frac{x^2}{4r_0}\,, \qquad x^5 = r_0\eta\,,
\ee
where $\eta$ is an angle, the metric (\ref{5m1}) can be rearranged to
\be\lb{5m2}
ds_5^2 = -\bigg(1 - \frac{\mu}{x^2}\bigg)\,dt^2 + \bigg(1 -
\frac{\mu}{x^2}\bigg)^{-1}\,dx^2 + x^2\,d\Omega_3^2\,,
\ee
where
\be\lb{S3}
d\Omega_3^2 = \frac14\left(d\theta^2 + \sin^2\theta\,d\varphi^2  +
(d\eta - \cos\theta\,d\varphi)^2\right)
\ee
is the metric of the three-sphere. We recognize in (\ref{5m2}) the
static, spherically symmetric (in the four spatial dimensions)
Tangherlini-Myers-Perry (TMP) five-dimensional black hole
\cite{Tang,MP5} with mass parameter
\be\lb{mub}
\mu = 4r_0b.
\ee
So the static NAF
magnetic black holes of \E3 are simply the dimensional reduction of
the asymptotically flat TMP black holes relative to one of the
azimuthal angles. Correspondingly, the NAF
extreme ($b=0$) magnetic black holes of \E3 are simply the
dimensional reduction of five-dimensional Minkowski spacetime, their
timelike singularity $r=0$ arising from the dimensional reduction of
the (spurious if $x^5$ is periodic with period $4\pi r_0$) Dirac
string singularity of (\ref{5m1}).

The reduced 3-metric $\gamma_{ij}$ of (\ref{5m1}) coincides with that
of the four-dimensional Schwarzschild solution with mass $M = b/2$.
It follows that there is an $SL(3,R)$ transformation $U_{Sm}$ which
transforms the five-dimensional trivially embedded Schwarzschild
solution, i.e. the direct product of the four-dimensional
Schwarzschild solution with the line (or the circle), with
representative matrix
\be\lb{chiS}
\chi_S = \left( \begin{array}{ccc} -\frac{r-b}{r} & 0 & 0 \\ 0 & 1 & 0
\\ 0 & 0 & -\frac{r}{r-b}  \end{array}
\right)\,,
\ee
into the TMP black hole, written in the form (\ref{5m1}), with
representative $\chi_m$:
\be
\chi_m = U_{mS}\chi_S U_{Sm}\,, \qquad U_{Sm} =\left(
\begin{array}{ccc}
1&0&0\\0&0&\sqrt{\frac{r_0}{b}}
\\ 0&-\sqrt{\frac{b}{r_0}}&\sqrt{\frac{r_0}{b}}
\end{array}
\right)
\ee
(with $U_{mS}=U_{Sm}^T$).

As explained above, this transformation acting
on the trivially embedded Kerr solution will generate rotating NAF
magnetic black holes according to
\be\lb{transfK}
\chi_{mK} = U_{mS}\chi_K U_{Sm}.
\ee
The trivially embedded Kerr solution with mass $M_0 = b/2$ and rotation
parameter $a_0$ is (in Boyer-Lindquist coordinates)
\ba\lb{K}
ds_5^2 & = & -\bigg(\frac{\Gamma_0}{\Sigma_0}\bigg)
\bigg(dt - \omega_0\,d\varphi\bigg)^2 \nonumber\\ && \qquad +
\Sigma_0\bigg(\frac{dr^2}{\Delta_0}+d\theta^2+\frac{\Delta_0\sin^2\theta}
{\Gamma_0}\,d\varphi^2 \bigg) + (dx^5)^2\,,
\ea
with
\ba \Delta_0
& = & r^2- br+a_0^2, \quad  \Sigma_0 =  r^2+a_0^2\cos^2\theta, \lb{DS}\\
\Gamma_0 & = & \Delta_0 - a_0^2\sin^2\theta\,, \quad \omega_0   = 
-\frac{a_0br\sin^2\theta}{\Gamma_0}\,.
\ea
Acting on its representative matrix
\be
\chi_K =\frac1{\Gamma_0}\left(
\begin{array}{ccc}
-(r-b)^2-a_0^2\cos^2\theta & 0 &
a_0b\cos\theta \\ 0  & \Gamma_0 & 0 \\
a_0b\cos\theta & 0 & -\Sigma_0
\end{array}
\right)
\ee
with the transformation (\ref{transfK}), we obtain
\be
\chi_{mK} =\frac1{\Gamma_0}\left(
\begin{array}{ccc}
-(r-b)^2-a_0^2\cos^2\theta &  -a_0b\sqrt{\frac{b}{r_0}}\cos\theta &
a_0\sqrt{br_0}\cos\theta \\  -a_0b\sqrt{\frac{b}{r_0}}\cos\theta  &
-\frac{b}{r_0}\Sigma_0 & \Sigma_0 \\ a_0\sqrt{br_0}\cos\theta
& \Sigma_0 & -r_0r
\end{array}
\right).
\ee
The resulting five-dimensional metric is
\ba
ds_5^2 & = & -\bigg(1-\frac{b}{r}\bigg)\psi^2
+\frac{a_0\sqrt{br_0}\cos\theta}{r} 2\psi\xi
+ \frac{r_0\Sigma_0}{r}\xi^2 \nonumber \\
&&+ \frac{r_0r}{\Gamma_0}\bigg(\frac{\Gamma_0}{\Delta_0}\,dr^2 +
\Gamma_0\,d\theta^2 + \Delta_0\sin^2\theta\,d\varphi^2
\bigg)\,, \lb{mrot5}
\ea
with
\be
\psi = dt+\frac{a_0\sqrt{br_0}r\sin^2\theta}{\Gamma_0}\,d\varphi\,,
\quad \xi = d\eta-\frac{\Delta_0\cos\theta}{\Gamma_0}\,d\varphi\,,
\ee
where as before we have put $x^5 = r_0\eta$. Not surprisingly,
this turns out to coincide with a subclass of
rotating Myers-Perry black holes \cite{MP5}. The general rotating
Myers-Perry (MP) black hole in five dimensions depends on two angular
momentum parameters $a_+$ and $a_-$ and is given by
\ba\lb{mp}
ds_5^2 & = & -dt^2 + \frac{\mu}{\rho^2}\bigg[dt+a_+\sin^2\tb\,d\varphi_+
+a_-\cos^2\tb\,d\varphi_-\bigg]^2 + \rho^2\,d\tb^2 \nonumber
\\ & & + (x^2+a_+^2)\sin^2\tb\,d\varphi_+^2 +
(x^2+a_-^2)\cos^2\tb\,d\varphi_-^2 + \frac{\rho^2x^2}{\Theta}\,dx^2\,.
\ea
with
\be
\rho^2 = x^2 + a_+^2\cos^2\tb + a_-^2\sin^2\tb\,, \quad
\Theta = (x^2 + a_+^2)(x^2 + a_-^2) - \mu x^2\,.
\ee
After transforming the spatial coordinates to
\be
\tb = \theta/2\,, \quad \varphi_{\pm} = (\varphi\pm\eta)/2\,,
\quad x^2 = 4r_0r - \bar{a}^2\,,
\ee
the metric (\ref{mrot5}) is found to go over into the MP metric with
two equal angular momentum parameters, the identification being
\be
\mu = 4r_0b\,,\quad a_+ = a_- = \bar{a} =
-\frac{4r_0a_0}{\sqrt{\mu}}\,.
\ee

The Kaluza-Klein dimensional reduction of (\ref{mrot5}) leads to the
rotating NAF magnetic black hole solution
\ba
ds_4^2 & = & -
\frac{\Gamma_0}{\sqrt{r_0r\Sigma_0}}\bigg(dt +
\frac{\sqrt{br_0}a_0r\sin^2\theta}
{\Gamma_0}\,d\varphi\bigg)^2 \nonumber \\ & &
+\sqrt{r_0r\Sigma_0} \bigg(\frac{dr^2}{\Delta_0} + d\theta^2 +
\frac{\Delta_0\sin^2\theta}{\Gamma_0}\,d\varphi^2
\bigg)\,, \lb{mrotg} \\ \A & = &
-\frac{r^2+a_0^2}{\Sigma_0}\frac{r_0}2\cos\theta\bigg(d\varphi -
\frac{a_0\sqrt{b}}{\sqrt{r_0}(r^2+a_0^2)}\,dt\bigg)\,, \lb{mrota} \\
\e^{2\phi/\sqrt3} & = & \sqrt{\frac{r_0r}{\Sigma_0}}\,.\lb{mrotp}
\ea

Similarly to the Kerr spacetime, the rotating spacetime (\ref{mrotg})
possesses two horizons at $r = r_{\pm}$, with
\be
r_{\pm} = \frac12\bigg(b \pm \sqrt{b^2-4a_0^2}\bigg)\,,
\ee
$r=r_+$ corresponding to the event horizon. However, to the
difference of the Kerr black hole, but similarly to the case of
rotating linear dilaton black holes ($\alpha =1$) \cite{newdil1}, the
metric (\ref{mrotg}) (as well as the five-dimensional metric
(\ref{mrot5})) is singular on the timelike line $r =0$.
Accordingly, the Penrose diagrams along the symmetry axis of these NAF 
rotating black holes
are, for the three cases  $b^2>4a_0^2$, $b^2=4a_0^2$ and $b^2<4a_0^2$,
identical to those of the Reissner-Nordstr\"{o}m spacetime, with the
charge replaced by the angular momentum parameter $a_0$. According to
(\ref{mrota}), this metric supports a rotating monopole
magnetic field with magnetic charge
\be\lb{P}
P = \frac1{4\pi}\int F_{\theta\varphi}\,d\theta\,d\varphi =
\frac12\,A_{\varphi}\bigg]^{\theta = 0}_{\theta = \pi} = -\frac{r_0}2\,,
\ee
as in the static case. The quasilocal computation of the mass $\M$ and
angular momentum $J$ of these black holes, discussed in the next
section, gives
\be
\M = \frac{3b}8\,, \qquad J = \frac{a_0}2\sqrt{br_0}\,.
\ee

In the massless case $M_0=b/2=0$, we know that the Kerr spacetime
reduces to Minkowski spacetime. Therefore the spatial part of the
five-dimensional metric (\ref{K}) is a (spheroidal coordinate)
parametrization of flat three-space. This means that the
solutions (\ref{mrotg})-(\ref{mrotp}) with $b=0$ can be linearly
superposed to give multicenter solutions similar to the
Israel-Wilson-Perj\`es \cite{iwp} (IWP) solutions of Einstein-Maxwell
theory. For $b = 0$, the matrix
$A_m$ of (\ref{mA}) is such that $A^2=0$, so that the target space
geodesic (\ref{targeo}) is null \cite{spat,spat2}. The corresponding
five-dimensional metric
\be\lb{miwp}
ds_5^2 = -dt^2 + d{\bf x}^2 +\sigma^{-1}(\,dx^5 +
\mbox{\LARGE$a$}_i^5\,dx^i)^2\,,
\ee
with
\be
\nabla\wedge\mbox{\LARGE$a$} = \nabla\sigma\,,
\ee
reduces to the magnetic equivalent of the four-dimensional solution
(\ref{nugeog})-(\ref{nugeoe}) with $\alpha= \sqrt{3}$ ($\gamma =
-1/2$). For the choice of the harmonic function
\be\lb{harsphe}
\sigma=\frac{r_0r}{r^2+a_0^2 \cos^2\theta}\,,
\ee
this coincides with the singular solution (\ref{mrotg})-(\ref{mrotp})
with $b=0$. As we have seen, for $a_0=0$ the corresponding
five-dimensional metric reduces to five-dimensional Minkowski if
$x^5$ is periodic with period $4\pi r_0$. Accordingly, the static
multi-center harmonic function (\ref{multi}) with $\sigma_{\infty}=0$ 
will lead to five-dimensional
spacetimes with Dirac string singularities originating from the
centers, unless all the residues $c_i$ are equal, in which case
(\ref{miwp}) will reduce again to five-dimensional Minkowski spacetime.

\setcounter{equation}{0}
\section{Case $\alpha^2 = 3$: dyonic black holes (magnetic sector)}

We have seen that the static NAF magnetic black holes of \E3
are a dimensional reduction of the five-dimensional TMP black holes,
while their rotating generalisations (\ref{mrotg})-(\ref{mrotp}) are a
dimensional reduction of the MP black holes with two equal angular
momentum parameters. One may surmise that the more general MP black
holes (\ref{mp}) with two unequal angular momentum parameters
should lead, by an appropriate dimensional reduction, to
generalized four-dimensional NAF magnetic black holes with two extra
quantum numbers, which as we will now show can be identified as
four-dimensional angular momentum and electric charge associated
with rotating dyonic black holes. Putting
\be\lb{apm}
a_{\pm} = \beta \pm \delta\,,
\ee
transforming as before the spatial coordinates to
\be
\tb = \theta/2\,, \quad \varphi_{\pm} = (\varphi\pm\eta)/2\,,
\quad x^2 = 4r_0r - (\beta^2+\delta^2)\,,
\ee
carrying out the dimensional reduction (\ref{4+1}) of
(\ref{mp}) relative to $\partial_5 = r_0^{-1}\partial_{\eta}$,
and putting $\mu = 4r_0b$, we obtain
\ba
ds_4^2 & = & - \frac{\Gamma}{\sqrt{\Pi A}}\bigg(dt - \bar{\omega}
\,d\varphi\bigg)^2 +\sqrt{\Pi A} \bigg(\frac{dr^2}{\Delta} +
d\theta^2 + \frac{\Delta\sin^2\theta}{\Gamma}\,d\varphi^2 \bigg)\,,
\lb{drotg} \\ \A & = & -\frac{r_0}{2A}\bigg(\bigg[(\Delta +
br)\cos\theta -\frac{b\beta\delta}{2r_0} -
\frac{\beta\delta}{2r_0}(r-b/2)\sin^2\theta\bigg]d\varphi \nonumber
\\ & & \qquad - \frac{b}{2r_0}(\delta-\beta\cos\theta)\,dt\bigg)\,,
\lb{drota} \\ \e^{2\phi/\sqrt3} & = &
\sqrt{\frac{\Pi}{A}}\,,\lb{drotp}
\ea
with
\ba\lb{Daf2}
\Delta & = & (r-b/2)^2 - \frac{\bgb^2\bgd^2}{4r_0^2}\,, \quad \Gamma
= \Delta - \frac{\beta^2\bgd^2}{4r_0^2}\,\sin^2\theta\,,  \nonumber
\\ A & = & r^2 + \frac{b}{4r_0}(\delta - \beta\cos\theta)^2 -
\frac{\beta^2\delta^2}{4r_0^2}\cos^2\theta\,, \quad \Pi = r_0r
+\frac{\beta\delta}2\cos\theta\,, \\ \bar{\omega} & = & \frac{\beta
b(r-\delta^2/2r_0)}{2\Gamma}\sin^2\theta \qquad
(\bgb^2=br_0-\beta^2\,,\; \bgd^2 = br_0 - \delta^2)\,. \nonumber
\ea
We see that the reduced spatial metric in
(\ref{drotg}) coincides with that of the Kerr metric with parameters
$M^2 = b\bgd^2/4r_0$, $a^2 = \beta^2\bgd^2/4r_0^2$, ,
meaning that the solutions (\ref{drotg})-(\ref{drotp}) can also be
generated from the trivially embedded Kerr metric, or from the
rotating magnetic solutions (\ref{mrotg})-(\ref{mrotp}), by a suitably
chosen $SL(3,R)$ transformation.

The spacetimes (\ref{drotg}) have two horizons $r = r_{\pm}$, with
\be
r_{\pm} = \frac12\bigg(b \pm \frac{\bgb\bgd}{r_0}\bigg),
\ee
if either $\beta^2 <
br_0$ and $\delta^2 < br_0$, or $\beta^2 > br_0$ and $\delta^2 >
br_0$. However, as discussed in \cite{MP5}, this second possibility
does not lead to regular black holes. The values of $A_{\varphi}$ on
the Dirac strings $\theta = 0$ and $\theta = \pi$ are unchanged from
their static ($\beta = 0$) values, so that the magnetic charge is
again given by (\ref{P}),
\be\lb{Pm}
P = -\frac{r_0}2.
\ee
In order to compute the electric charge, we evaluate
\ba
\Pi^r & \equiv & \frac1{4\pi}\sqrt{|g|}\e^{-2\sqrt3\phi}F^{rt}
\nonumber \\
& = & \frac{\sin\theta}{8\pi}\frac{br_0}{\Pi^2}\bigg(
(\delta-\beta\cos\theta)r^2 -
\frac{\beta\delta}{4r_0}[\beta(1+\cos^2\theta)
-2\delta\cos\theta]r \nonumber \\ & & \quad
-\frac{\beta^2\delta^3}{8r_0^2}\sin^2\theta\bigg),
\ea
leading after integration to
\be\lb{Qm}
Q = -\frac{b\delta}{2r_0}.
\ee
So the difference $\delta$ between the two angular momentum
parameters of the original MP black hole (\ref{mp}) leads, after
dimensional reduction, to a four-dimensional electric charge, while
the sum $\beta$ leads to a four-dimensional angular momentum, which
shall be computed below.

For $\delta=0$ ($a_- = a_+$), we recover the rotating magnetic black
holes of the previous section, with
\be
a_0 = -\sqrt{b/r_0}\,\beta/2\,.
\ee
For $\beta = 0$ ($a_- = -a_+$), the solution takes the form
\ba
ds_4^2 & = &
-\frac{(r-b/2)^2-b\bgd^2/4r_0}{\sqrt{r_0r(r^2+b\delta^2/4r_0)}}\,dt^2
\nonumber\\&&
+ \sqrt{r_0r(r^2+b\delta^2/4r_0)}
\bigg(\frac{dr^2}{(r-b/2)^2-b\bar{\delta}^2/4r_0} + d\Omega^2\bigg),
\lb{dsg}\\ \A & = & -\frac{r_0}{2}\cos\theta\,d\varphi
+\frac{b\delta}{4(r^2+b\delta^2/4r_0)}\,dt, \lb{dse}\\
\e^{2\phi/\sqrt3} & = & \sqrt{\frac{r_0r}{r^2+b\delta^2/4r_0}},\lb{dsp}
\ea
corresponding to static NAF dyonic black holes. Finally, for $b = 0$ 
we recover the magnetostatic IWP solutions discussed at
the end of the previous section. 

The NAF black holes (\ref{drotg}) become extreme for $\bgb = 0$
($\beta^2 = br_0$) or $\bgd = 0$ ($\delta^2 = br_0$), In this last
case ($\bgd = 0$), it follows from 
(\ref{Daf2}) that $\Delta = \Gamma = (r-b/2)^2$, so
that the solutions (\ref{drotg}) have conformally flat spatial
sections. Again, this means that these solutions can be linearly 
superposed to give multicenter solutions. We only discuss here the
static case ($\beta=0$, $\delta^2 = br_0$, i.e. $a_+ = -a_-
= \sqrt{\mu/4}$). In this case the solution (\ref{dsg})-(\ref{dsp}) is
of the form (\ref{targeo}), with the matrix $\eta$ given in
(\ref{mA}),
\be
A =\left(
\begin{array}{ccc}
-1&-\sqrt{\frac{b}{r_0}}&0\\ 0&1&-\frac{r_0}b\\ \sqrt{\frac{b}{r_0}}&
\frac{b}{r_0}&0
\end{array}
\right)\,,
\ee 
and $\sigma = b/(r-b/2)$. This matrix $A$ is such that $A^2 \neq 0$,
$A^3 = 0$. The corresponding multicenter solution thus 
belongs to the class (b) discussed in \cite{spat} and is given by
\ba
ds_4^2 & = & -fdt^2 + f^{-1}d{\bf x}^2, \quad
f = \sqrt{\frac{b}{r_0}}
\sqrt{\sigma(1+\sigma/2)(1+\sigma+\sigma^2/2)}\;\;\;\; \lb{iwpdg}\\ 
\A & = & \frac{r_0}{2b}\,\sA_i\,dx^i + \sqrt{\frac{r_0}{b}}
\frac{\sigma^2}{4(1+\sigma+\sigma^2/2)}\,dt\,, \lb{iwpdma}\\
\e^{2\phi/\sqrt3} & =&
  \sqrt{\frac{r_0}{b}}\sqrt{\frac{\sigma(1+\sigma/2)}
{1+\sigma+\sigma^2/2}}\,, \lb{iwpdmp}
\ea
with
\be\lb{As}
\sigma = \Sigma_i(c_i/r_i)\,, \quad \nabla\wedge\sA = \nabla\sigma\,.
\ee

Now we compute the mass and angular momentum of the four-dimensional
$\alpha = \sqrt{3}$ rotating dyonic NAF black holes. The mass is again given
by the limit $r \to \infty$ of the difference between the values of
the quasilocal Hamiltonian (\ref{ham}) for the rotating ($r_0$, $b$, $\beta$,
$\delta$) black hole and for the ($r_0$, $0$, $0$, $0$) background. The ADM
form of the metric (\ref{drotg}) is
\ba
ds^2 & = & - \frac{\sqrt{\Pi A}\,\Delta}{\Omega}\,dt^2 +
\sqrt{\Pi A}\bigg[ \frac{dr^2}{\Delta} + d\theta^2 \nonumber \\ && +
\frac{\Omega}{\Pi A}\,\sin^2\theta\bigg(d\varphi + \frac{\beta
b(r-\delta^2/2r_0)}{2\Omega}\,dt\bigg)^2\bigg],
\ea
with
\be
\Omega = \Pi\Delta + br_0(r^2-\beta^2\delta^2/4r_0^2)\,.
\ee
Keeping only the leading asymptotic ($r \to \infty$) monopole and
dipole contributions, we obtain
\be
\sigma_{ab}dx^adx^b \simeq r\sqrt{\Pi}d\Omega^2, \quad n^r \simeq N
\simeq \frac{\sqrt{r-b}}{\Pi^{1/4}}, \quad N^{\varphi} \simeq
\frac{\beta b}{2r_0r^2},
\ee
leading to the asymptotic extrinsic curvature
\be
k \simeq -\frac32r_0^{-1/4}r^{-3/4}\bigg(1-\frac{b}{2r} -
\frac7{24}\frac{\beta\delta}{r_0r}\cos\theta\bigg) .
\ee
So the value of the first term $N\epsilon$ of the integrand of
(\ref{ham}) differs from the magnetostatic ($\beta=\delta=0$) value
only by a term in $\sin\theta\cos\theta$ which disappears after
integration over the sphere. The second term of the integrand behaves
asymptotically as the product of $N^{\varphi}$, in $r^{-2}$,  with
\be\lb{gravspin}
\pi_{\varphi r}n^r \simeq \frac{b\beta}{32\pi}\sin^3\theta\,,
\ee
so that its contribution to the quasilocal mass vanishes when the
radius of the sphere $S^t_r$ is taken to infinity. Finally, the
third term of the integrand behaves asymptotically as the product of
\be
A_0 \simeq \frac{b}{4r_0r^2}(\delta-\beta\cos\theta)
\ee
with
\be
\Pi^r \simeq \frac{b\sin\theta}{8\pi r_0}(\delta-\beta\cos\theta),
\ee
and again goes to zero when $r \to \infty$. It follows
that the mass of the rotating dyonic black hole is unchanged from
its magnetostatic value,
\be
\M = \frac{3b}8\,.
\ee

The quasilocal momenta are obtained from the Hamiltonian by carrying
out an infinitesimal gauge transformation $\delta x^i =
\delta\xi^i\,t$ and evaluating the response $P_i = \delta
H/\delta\xi^i$. For the angular momentum $J = P_{\varphi}$, one
obtains \cite{HaRo95}
\be\lb{J}
J = -\oint_{S_t^{r}}(2\pi_{\varphi r}n^r + A_{\varphi}\Pi^r)
\ee
(the static background does not contribute). From the preceding
evaluations and
\be
A_{\varphi} \simeq -\frac{r_0}2\cos\theta,
\ee
one obtains asymptotically
\be\lb{emspin}
A_{\varphi}\Pi^r \simeq
-\frac{b\sin\theta}{16\pi}(\delta\cos\theta-\beta\cos^2\theta).
\ee
The first term disappears after integration over the sphere, while the
second term combines with the gravitational contribution
(\ref{gravspin}) to yield
\be\lb{Jm}
J = -\frac{b\beta}{4}\,.
\ee
It is remarkable that, as in the case of the linear dilaton black
holes \cite{newdil1}, the (often ignored) electromagnetic
contribution to the angular momentum is of the same order as the
purely gravitational contribution.

There is a simple relation between these values of the mass and
angular momentum for the four-dimensional NAF black holes
(\ref{drotg})-(\ref{drotp}) and the values of the corresponding
quantities \cite{MP5} for the five-dimensional Myers-Perry black holes
(\ref{mp})
\be
M_5 = \frac{3\pi}8\,\mu\,, \qquad J_{5\pm} = \frac23\,M_5a_{\pm}\,,
\ee
which may be understood in terms of the dimensional
reduction procedure. This amounts to integrating out the cyclic 
coordinate $x^5 = r_0\eta$ from the five-dimensional action. The
perimeter $\oint dx^5$ may be evaluated by computing the area $A_3$ 
($=2\pi^2$) of the three-sphere from (\ref{S3}),
\be
A_3 = \bigg(\frac14\bigg)^{3/2}\,A_2\oint d\eta \,.
\ee
Using the relations (\ref{mub}) and (\ref{apm}), we then check that the
ratio between the four- and five-dimensional parameters is equal to 
the perimeter of the cyclic dimension:
\be
\frac{M_5}{M_4} = -\frac{J_{5+}+J_{5-}}{2J_4} =
\frac{J_{5+}-J_{5-}}{2QP} = \oint dx^5 = 4\pi r_0\,,
\ee

We now check that these values of the black hole mass and angular
momentum, together with the other physical parameters of the rotating
black holes (\ref{drotg})-(\ref{drotp}), satisfy the generalized first law of
thermodynamics for dyonic, rotating black holes \cite{Rasheed},
\be\lb{firstdyon}
d\M =TdS + \Omega_h dJ + V_h dQ + W_h dP\,.
\ee
In (\ref{firstdyon}),
the Hawking temperature and entropy are
\be
T = \frac{\bgb\bgd}{2\pi br_0(\bgb+\bgd)}\,, \quad S =
\frac{\pi b}2\,(\bgb + \bgd)\,.
\ee
$\Omega_h$ is the horizon angular velocity
\be
\Omega_h =  - N^{\varphi}\big|_h = -\frac{\beta\bgd}{br_0(\bgb+\bgd)}\,.
\ee
$V_h$ is the proper horizon electric potential, i.e.
the electric potential in the horizon rest frame \footnote{The
electric potential in (\ref{drota}) is uniquely defined by the
condition that it vanish at spatial infinity.}, evaluated on the
horizon $r=r_+$,
\be
V_h = -\big(A_0 - N^{\varphi}A_{\varphi}\big)_h =
-\frac{\delta\bgb}{2b(\bgb+\bgd)}\,,
\ee
while $W_h$ is the proper horizon magnetic potential, which (in the
absence of a better definition) we shall define by duality as the
proper horizon electric potential of the dyonic black hole of the
electric type (see Section 7) with electric charge $P$, magnetic
charge $Q$, and angular momentum $-J$ \cite {Rasheed}, given in (\ref{Wh}).
A straightforward computation then shows that when the three parameters
$b$, $\bgb$, $\bgd$ are varied independently, the
generalized first law (\ref{firstdyon}) is satisfied identically,
provided the scale parameter $r_0$ is held fixed. 

One also easily checks that the rotating dyonic black holes satisfy 
the Smarr-like formula
\be\lb{smarrdyon}
\M = \frac34\bigg(2TS + 2\Omega_hJ + V_hQ + W_hP\bigg)\,,
\ee
similar to the Smarr formula for the asymptotically flat dyonic black
holes of Kaluza-Klein theory \cite{Rasheed}, except for the anomalous
factor $3/4$. This factor may be understood by comparing
(\ref{smarrdyon}) (where $V_hQ+ W_hP = 2V_hQ$ on account of
(\ref{QVPW})) with the 5-dimensional Myers-Perry Smarr formula \cite{MP5}
\be
M_5 = \frac34\bigg(2TS_5 + 2\Omega_+J_+ + 2\Omega_-J_-\bigg)\,,
\ee 
where the 5-dimensional entropy (the quarter of the 5-dimensional
horizon area) is $S_5 = 4\pi r_0S_4$, and
$\Omega_{\pm}$ are the two horizon angular velocities.

\setcounter{equation}{0}
\section{Case $\alpha^2 = 3$: dyonic black holes (electric sector)}
From these NAF rotating dyonic black holes of the magnetic type, one
can construct a dual sector of NAF rotating dyonic black holes of the
electric type by the electromagnetic duality transformation
(\ref{emdual}). Before doing this, it is instructive to first
consider the purely electrostatic NAF black holes and their rotating
counterparts. For $\alpha = \sqrt{3}$, we obtain for our electric NAF
black hole solution (\ref{nbhg})-(\ref{nbhe})
\be
\e^{2\phi/\sqrt3} = \bigg(\frac{r}{r_0}\bigg)^{1/2}, \quad A_0 =
\frac{r-b}{2r_0},
\ee
leading for the 5-dimensional metric (\ref{4+1}) to
\be\lb{5e1}
ds_5^2 = -\frac{r-b}{r_0}\,dt^2 + \frac{r_0}{r}\bigg(dx^5 +
\frac{r-b}{r_0}\,dt\bigg)^2 + \frac{r}{r-b}\bigg(dr^2 +
r(r-b)\,d\Omega^2\bigg)\,.
\ee
The corresponding representative matrix $\chi_e$
\be\lb{chie}
\chi_e = \left( \begin{array}{ccc} -\frac{b(r-b)}{r_0r} &
\frac{r-b}{r}& 0
\\\frac{r-b}{r}  & \frac{r_0}{r} & 0 \\ 0 & 0 & -\frac{r}{r-b}  \end{array}
\right)
\ee
is of the form (\ref {targeo}) with the harmonic potential
(\ref{harm}) and the constant matrices
\be
\eta_e =\left(
\begin{array}{ccc}
-\frac{b}{r_0}&1&0\\ 1&0&0\\ 0&0&-1
\end{array}
\right), \quad A_e=\left(
\begin{array}{ccc}
-\frac{b}{r_0}&1&0\\ 0&0&0\\ 0&0&\frac{b}{r_0}
\end{array}
\right)\,. \label{eA}
\end{equation}

Again, the reduced 3-metric $\gamma_{ij}$ in (\ref{5e1}) coincides with that
of the Schwarzschild solution with mass $M = b/2$, meaning that the
matrix $\chi_e$ is an $SL(3,R)$ transform of the matrix (\ref{chiS})
for the trivially embedded Schwarzschild solution,
\be
\chi_e = U_{eS}\chi_S U_{Se}\,, \qquad U_{Se} =\left(
\begin{array}{ccc}
\sqrt{\frac{b}{r_0}}&-\sqrt{\frac{r_0}{b}}&0\\
0&\sqrt{\frac{r_0}{b}}&0\\ 0&0&1
\end{array}
\right)\,.
\ee
Equivalently, one can simply note that the 5-dimensional metric
(\ref{5e1}) can be rearranged as
\be\lb{ts2}
ds_5^2 = -\frac{b}{r_0}\bigg(1-\frac{b}{r}\bigg)\bigg(dt -
\frac{r_0}{b}\,dx^5\bigg)^2 + \frac{r}{r-b}\,dr^2 + r^2\,d\Omega^2 +
\frac{r_0}{b}(dx^5)^2,
\ee
which is clearly the product of the Schwarzschild solution for an
observer with time
\be\lb{twist}
t' = \sqrt{\frac{b}{r_0}}\bigg(t - \frac{r_0}{b}x^5\bigg)
\ee
with the line or the circle. So the NAF electric black hole solution
(\ref{nbhg})-(\ref{nbhe}) with $\alpha=\sqrt{3}$ is simply a twisted
dimensional reduction of the trivial 5-dimensional embedding of the
Schwarzschild solution with respect to the Killing vector
\be\lb{infbst}
\partial_5' =
\sqrt{\frac{r_0}b}\bigg(\partial_5-\partial_t\bigg)\,.
\ee
Note that the ``twist''\footnote{This term is used here in a
different sense from that used in (\ref{dual}).} in the relation
(\ref{twist}) between the two times is not innocent in the
traditional Kaluza-Klein interpretation, where the fifth coordinate
$x^5$ varies on the so-called Klein circle so that, assuming the
domain of the time $t$ to be the real axis, the corresponding
Schwarzschild time $t'$ must be periodically identified\footnote {A
similar phenomenon occurs for the spinning point particle solution of
2+1 gravity \cite{DJH} which is obtained from the non-spinning
solution $ds_3^2 = -dt^2 + dr^2 + \alpha^2r^2\,d\theta^2$ by the
replacement $t \to t + \omega\theta$ ($\alpha$, $\omega$ constant).}.

Let us recall that the well-known asymptotically flat electrically
charged Kaluza-Klein
black holes, given by (\ref{ghs}) with $\alpha=\sqrt{3}$, may also be
obtained from the trivially embedded Schwarzschild solution by a
twisted dimensional reduction, this time with respect to the boosted
Killing vector
\be\lb{bst}
\partial_5'' = \cosh\xi\,\partial_5 + \sinh\xi\,\partial_t
\ee
(the
$SL(3,R)$ transformations preserving staticity and asymptotic
flatness belong to the subgroup $SO(1,1)$ of Lorentz boosts in the
2-space $(t,x^5)$). The product of this boost with the transformation
(\ref{infbst}) --- an infinite boost together with an unessential
rescaling --- being again a rescaled infinite boost, it follows that
the twisted dimensional reduction with respect to (\ref{infbst}) of
the AF electrostatic Kaluza-Klein black holes will also lead to
essentially the same (up to a rescaling of $r_0$) NAF electrostatic
black hole\footnote{In this case there is no simple relation between the
electric charge of the AF black hole and that of the NAF black hole.}.

We are now in a position to generate the rotating electric NAF black
hole solutions of \E3. We have noted that the transformation from the
Schwarzschild solution of \E3 to the static solution
(\ref{nbhg})-(\ref{nbhe}) boils down in this case to the ``twist in
time'' (\ref{twist}). Therefore the rotating solutions are simply the
twisted dimensional reduction of the trivially embedded Kerr
solutions with mass $b/2$ and rotation parameter $a_0$, i.e. their
5-dimensional metric is (in Boyer-Lindquist coordinates)
\ba\lb{twistk}
ds_5^2 & = & -\frac{b}{r_0}\bigg(\frac{\Gamma_0}{\Sigma_0}\bigg)
\bigg(dt - \sqrt{\frac{r_0}b}\,\omega_0\,d\varphi - \frac{r_0}{
b}\,dx^5 \bigg)^2  \nonumber \\ && +
\Sigma_0\bigg(\frac{dr^2}{\Delta_0}+d\theta^2+\frac{\Delta_0\sin^2\theta}
{\Gamma_0}\,d\varphi^2 \bigg) + \frac{r_0}{b}\bigg(dx^5\bigg)^2\,,
\ea
Reversing the steps which led from the static solutions
(\ref{nbhg})-(\ref{nbhe}) with $\alpha = \sqrt{3}$ to (\ref{ts2}), we
obtain the rotating NAF electric black holes of \E3
\ba
ds_4^2 & = & - \frac{\Gamma_0}{\sqrt{r_0r\Sigma_0}}\bigg(dt +
\frac{\sqrt{br_0}a_0r\sin^2\theta} {\Gamma_0}\,d\varphi\bigg)^2
\nonumber \\ & & +\sqrt{r_0r\Sigma_0} \bigg(\frac{dr^2}{\Delta_0} +
d\theta^2 + \frac{\Delta_0\sin^2\theta}{\Gamma_0}\,d\varphi^2
\bigg)\,, \lb{erotg} \\ \A & = & \frac{1}{2r_0r}\bigg(\Gamma_0\,dt
+ \sqrt{br_0}a_0r\sin^2\theta\,d\varphi\bigg), \lb{erota}
\\ \e^{2\phi/\sqrt3} & = & \sqrt{\frac{\Sigma_0}{r_0r}}. \lb{erotp}
\ea
The corresponding electric field
\be\lb{erotF}
F  =  \frac{r^2-a_0^2\cos^2\theta}{2r_0r^2}\,dr\wedge dt +
\sqrt{\frac{b}{r_0}}a_0\cos\theta\sin\theta\,d\theta\wedge\bigg(d\varphi
- \frac{a_0}{\sqrt{br_0}r}\,dt\bigg)
\ee
leads again to the electric charge (\ref{Q}),
\be
Q = \frac{r_0}2.
\ee

As in the magnetic case, for $b=0$ the reduced three-dimensional
metric in (\ref{erotg}) is flat, so that the solution
(\ref{erotg})-(\ref{erotp}) with $b=0$ can be generalized to
multicenter solutions. Using the null geodesic construction with
(\ref{eA}) for $b=0$ as input, it is easy to show that the
corresponding 5-dimensional metric is \cite{Gi82} 
\be
ds_5^2 = 2\,dt\,dx^5 + \sigma(dx^5)^2 + d{\bf x}^2,
\ee
in accordance with (\ref{nugeog})-(\ref{nugeoe}) for $\alpha =
\sqrt3$. The solution (\ref{erotg})-(\ref{erotp}) with $b=0$ is
recovered for the choice (\ref{harsphe}) of the harmonic function $\sigma$.

Clearly, the rotating electric solutions (\ref{erotg})-(\ref{erotp})
are dual to the rotating magnetic solutions
(\ref{mrotg})-(\ref{mrotp}). More generally, acting on the rotating
dyonic solutions of the magnetic type (\ref{drotg})-(\ref{drotp})
with the duality transformation (\ref{emdual}) with the lower sign,
we obtain the new sector of NAF rotating dyonic black hole solutions:
\ba
ds_4^2 & = & - \frac{\Gamma}{\sqrt{\Pi A}}\bigg(dt - \bar{\omega}
\,d\varphi\bigg)^2 +\sqrt{\Pi A} \bigg(\frac{dr^2}{\Delta} +
d\theta^2 + \frac{\Delta\sin^2\theta}{\Gamma}\,d\varphi^2 \bigg)\,,
\lb{derotg} \\ \A & = &
\frac{1}{2\Pi}\bigg(\bigg[A-b(r+\delta^2/2r_0)\bigg]dt
-\frac{b}{r_0}\bigg[\delta\Pi\cos\theta \nonumber
\\ & & \qquad + \frac{\beta r_0}{2} (r +
\delta^2/2r_0)\sin^2\theta\bigg]d\varphi\bigg), \lb{derota}
\\ \e^{2\phi/\sqrt3} & = & \sqrt{\frac{A}{\Pi}}\,,\lb{derotp}
\ea
(the functions $\Delta$, $\Gamma$, $A$, $\Pi$, $\bar{\omega}$ are
given in (\ref{Daf2})) with electric and magnetic charges
\be\lb{QPe}
Q = \frac{r_0}{2}\,, \qquad P = -\frac{b\delta}{2r_0}\,.
\ee
Lifting these solutions to five dimensions according to (\ref{4+1}),
and carrying out the inverse ``twist'' transformation
\be
t = \sqrt{\frac{r_0}b}\,(t' + x'^{5})\,, \quad x^5 =
\sqrt{\frac{b}{r_0}}x'^{5}\,,
\ee
we arrive after some calculations to the asymptotically flat
five-dimensional metric
\ba\lb{ranu}
ds_5^2 &=& -\frac{\Gamma}{B}\bigg(dt' - \omega\,d\varphi \bigg)^2 
+ A\bigg(\frac{dr^2}{\Delta} + d\theta^2
+ \frac{\Delta\sin^2\theta}{\Gamma}\,d\varphi^2 \bigg)
\nonumber \\ && \qquad
+ \frac{B}{A}\bigg(dx'^{5} + 2\hat{A}_{\mu}\,dx'^{\mu}\bigg)^2 
\ea
with
\ba
B & = & \bigg(r-\frac{\delta^2}{2r_0}\bigg)^2 +
\frac{\bgd^2}{4r_0^2}\bigg(\delta - \beta\cos\theta\bigg)^2 \,, 
\\ \omega &=& -\sqrt{\frac{b}{r_0}}\,
\frac{\delta\Delta\cos\theta - (\beta\bgd^2/2r_0)r\sin^2\theta}{\Gamma}\,,
\\ \hat{A}_0 &=&
-\frac{\delta}{2r_0}\,\frac{\delta(r-b/2)+(\beta\bgd^2/2r_0)\cos\theta}{B}
\,, \\
\hat{A}_{\varphi} & = & -\frac{\delta}2\,\sqrt{\frac{b}{r_0}}\,
\frac{[r^2 - (br_0\delta^2-\beta^2\bgd^2)/4r_0^2]\cos\theta 
+ (\beta\delta/2r_0)(r-b/2)\sin^2\theta}B\,.
\ea

The asymptotically flat metric (\ref{ranu}) describes rotating dyonic 
black holes with NUT charge, generalizing the five-dimensional
rotating dyonic black holes given by Rasheed \cite{Rasheed}. The
corresponding mass, scalar charge, NUT charge, electric charge, 
magnetic charge, and angular momentum are given by
\ba
M_{AF} & = & \frac{b}2 - \frac{\delta^2}{4r_0}\,, \quad \Sigma_{AF} = 
- \frac{\delta^2\sqrt{3}}{4r_0}\,, \quad N_{AF} =
\frac{\delta}2\,\sqrt{\frac{b}{r_0}}\,, \nonumber \\
Q_{AF} & = & - \frac{\delta^2}{2r_0}\,, \quad P_{AF} = -\frac{\delta}2\,
\sqrt{\frac{b}{r_0}}\,, \\
J_{AF} & = &
\frac{\beta\bgd^2}{4r_0}\,\sqrt{\frac{b}{r_0}}\,. \nonumber
\ea
There is no simple relation between these and the corresponding charges
for the NAF dyonic black hole solutions
(\ref{derotg})-(\ref{derotp}). The appearance here of an AF NUT charge
equal in magnitude to the AF magnetic charge is easy to
understand. Generically, the twist transformation (\ref{infbst}) acting on a
magnetized AF solution will  generate a NUTted NAF
solution, the NAF NUT charge vanishing only if $N_{AF} +
P_{AF}= 0$. Recalling that the combination of any
$SO(1,1)$ boost (\ref{bst}) with the twist acting on an AF Kaluza-Klein 
black hole will lead to essentially the same $\xi = 0$ NAF black hole
solution, we conclude that that the twist transformation acting on any
NUTted dyonic AF black hole with $N_{AF} + P_{AF} = 0$ will generate a
NUTless dyonic NAF black hole of the electric type 
(\ref{derotg})-(\ref{derotp}), which as we have seen is equivalent by
electromagnetic duality to the dimensional reduction of a Myers-Perry
black hole (\ref{mp}) relative to $\partial_{\eta} =
\partial_{\varphi_+} - \partial_{\varphi_-}$. More generally, we 
conjecture that the twist transformation acting on a generic
NUTted dyonic AF black hole will generate a
{\em NUTted} dyonic NAF black hole, equivalent by
electromagnetic duality to the dimensional reduction of a {\em
twisted} Myers-Perry black hole (\ref{mp}). This conjecture is proven
in the Appendix in the case where the original five-dimensional
solution is the trivially embedded Taub-NUT solution. 

Again, as in the magnetic case, for $\bgd = 0$ ($\delta^2 = br_0$) 
these dyonic black hole
solutions become extreme and can be linearly superposed to give
multicenter IWP solutions. Remarkably, the twisted asymptotically flat
metric (\ref{ranu}) reduces in this case to the static, pure NUT form
\ba
ds_5^2 &=& -(dt' + b\cos\theta d\varphi)^2 + \frac{A}{(r-b/2)^2}\bigg(
dr^2 +(r-b/2)^2 d\Omega^2\bigg)\nonumber \\
&& \quad + \frac{(r-b/2)^2}A\bigg(dx'^5 + 2
\hat{A}_{\mu}dx'^{\mu}\bigg)^2, \lb{ranue}
\ea
with $J_{AF} = 0$, so that the parameter $\beta$ seems to be
irrelevant. In the case $\beta = 0$, the solution 
(\ref{ranue}) is of the form (\ref{targeo}) with $\eta =$ diag(-1,1,-1),
\be
A =\left(
\begin{array}{rrr}
0&1&1\\ -1&-1&-1\\1&1&1
\end{array}
\right)\,,
\ee 
and $\sigma = b/(r-b/2)$. Again, this matrix $A$ is such that $A^2 \neq 0$,
$A^3 = 0$. After carrying out the twist transformation, we arrive at
the NAF multi-extreme black hole solution given by (\ref{iwpdg}) and
\ba
\A & = & \frac{b}{2r_0}\frac{1-\sigma^2/2}{\sigma(1+\sigma/2)}\,dt
+\frac12\sqrt{\frac{b}{r_0}}\sA_i\,dx^i\,, \lb{iwpdea}\\
\e^{2\phi/\sqrt3} & =&
  \sqrt{\frac{b}{r_0}}\sqrt{\frac{1+\sigma+\sigma^2/2}
{\sigma(1+\sigma/2)}}\,, \lb{iwpdep}
\ea 
where $\sigma$ and $\sA$ are given by (\ref{As}).

Let us now compute the quasilocal mass and angular momentum of
the ``electric'' dyonic black holes (\ref{derotg})-(\ref{derotp}). The 
spacetime metric is the same as for the ``magnetic'' dyonic black
holes, so that the gravitational contributions to the quasilocal energy
and angular momentum are the same. To compute the electromagnetic
contributions, we evaluate the asymptotic behavior of the electric
field density
\ba
\Pi_{(e)}^r & \equiv & \frac1{4\pi}\sqrt{|g|}\e^{-2\sqrt3\phi}F^{rt} = 
-\frac1{4\pi}F_{(m)\theta\varphi} \nonumber \\
& \simeq -& \frac1{8\pi}(r_0\sin\theta + \beta\delta\sin\theta\cos\theta)\,,
\ea
where $F_m$ is the dual electromagnetic field derived from
(\ref{drota}). The electromagnetic contribution to the quasilocal
mass is the integral over a large sphere of the product of 
\be
A_0 \simeq \frac{r}{2r_0}
\ee
(a fixed boundary data) with the substracted electric field density
\be
\Pi^r - \Pi^r|_0 \simeq -\frac{\beta\delta}{8\pi r}\sin\theta\cos\theta   
\ee
(where the index 0 stands for the static background ($b = \beta =
\delta = 0$)). This product goes to the finite value
$-(\beta\delta/16\pi r_0)\sin\theta\cos\theta$, which however averages
to zero after integration over the sphere, so that the mass is the
same as in the magnetic case. The electromagnetic contribution to
the quasilocal angular momentum is the integral of 
\be
A_{\varphi}\Pi^r \simeq
\frac{b\sin\theta}{16\pi}(\delta\cos\theta+\beta\sin^2\theta)
\ee
(compare with (\ref{emspin})). Again the first term disappears after
averaging over the sphere, while the integral of the second term is the
same as in the magnetic case. So the mass and angular momentum of
these ``electric'' dyonic black holes are, not surprisingly, the same 
as those of their dual magnetic counterparts
\be\lb{MJe}
\M = \frac{3b}8\,, \quad J = -\frac{b\beta}4\,.
\ee

Finally we return to the question of the validity of the generalized
first law (\ref{firstdyon}) for rotating dyonic black holes. For a
black hole of the magnetic sector (index $(m)$), the potential
$W_{(m)h}$ is the proper horizon electric potential for the dyonic
black hole of the electric sector (index $(e)$) with electric and
magnetic charges exchanged, and opposite angular momentum,
i.e. (compare (\ref{Qm}), (\ref{Pm}), (\ref{Jm}) on the one hand 
and (\ref{QPe}), (\ref{MJe}) on the other hand) the charge and parity 
conjugate ($A_0 \to -A_0$,
$N^{\varphi} \to - N^{\varphi}$) of the dyonic black hole
(\ref{derotg})-(\ref{derotp}) . We obtain for this quantity  the value
\be\lb{Wh} W_{(m)h} = -V_{(e)h} = \big(A_{(e)0} -
N_{(e)}^{\varphi}A_{(e)\varphi}\big)_h =
-\frac{\delta^2\bgb}{2r_0^2(\bgb+\bgd)}\,, \ee    which was used in
Sect. 6 to check the generalized first law (\ref{firstdyon}) in the
magnetic sector. Symmetrically, it follows  from this definition of
the proper horizon magnetic potential that $W_{(e)h} = V_{(m)h}$,
leading to the conclusion that the generalized first law is also
satisfied (as it should by duality) in the electric sector. Another
noteworthy consequence of (\ref{Wh}) is that the electric and magnetic
contributions to the generalized dyonic Smarr formula
(\ref{smarrdyon}) are equal, i.e. for dyons of either sector
\be\lb{QVPW} QV_h = PW_h = \frac{\delta^2\bgb}{4r_0(\bgb+\bgd)}\,.  \ee

\setcounter{equation}{0}
\section{Conclusion}

We have shown that the non-asymptotically flat static black hole
solutions of EMD theory may be obtained as near-horizon limits of 
asymptotically flat black holes, and recalled how their mass and
angular momentum may be computed unambiguously in the quasilocal
energy framework. 

We have then concentrated on the special case with dilaton coupling
constant $\alpha^2 = 3$, which is a dimensional reduction of
five-dimensional Kaluza-Klein theory. We have shown that in this case
the action of the group $SL(3,R)$ of invariance transformations may be
used to generate from the NAF magnetostatic or  electrostatic black
hole solutions two classes of NAF rotating dyonic black hole
solutions. The NAF black holes of the magnetic class turn out to be
simply dimensional reductions of the AF five-dimensional  Myers-Perry
black holes relative to one of the azimuthal angles, and their
four-dimensional quasi-local mass, quasi-local angular momentum, and
``dyonic momentum'' (the product of their electric and magnetic
charges) may be obtained from the mass and the two angular momenta of
the parent five-dimensional black hole by dimensional reduction. We
view this as a quite non-trivial success of the quasi-local energy
approach, insofar as the computation of the quasilocal mass and
angular momentum is rather involved.

On the other hand, the black holes of the electric class are unusual,
twisted dimensional reductions of the AF five-dimensional Rasheed
black holes with a NUT charge balancing the magnetic charge. It is
important to note that the extension of the local ``twist'' coordinate
transformation to a global coordinate transformation introduces a
periodicity in time if the fifth dimension is compactified on the
Klein circle, so that the NAF black holes of the electric sector are
only locally, not globally, equivalent to AF Rasheed black holes. A
by-product of this analysis is that, by electro-magnetic duality, there
is a one-to-one local correspondence between the Myers-Perry and
Rasheed classes of AF five-dimensional black holes.

We have also shown that the first law of black hole thermodynamics, as
generalized to dyonic black holes by Rasheed, is satisfied by both
classes of NAF rotating dyonic black holes, provided the overall scale
parameter $r_0$ is not varied. Again, this validity of the first law
under variation of three independent parameters is by no means
trivial, and constitutes another test of the validity of our
quasilocal energy computations. The NAF dyonic black holes are also
found to obey a generalized Smarr-like formula, Eq. (\ref{smarrdyon}).  

Finally, we have discussed the construction of NAF multi-extreme black
hole solutions. In the static case with generic $\alpha$, these are
actually singular multi-center solutions. In the case $\alpha^2 = 3$,
we have shown that, besides these, there are also two classes
(magnetic and electric) of regular, intrinsically dyonic multi-extreme
black hole solutions given by (\ref{iwpdg}) and either
(\ref{iwpdma})--(\ref{iwpdmp}) or (\ref{iwpdea})--(\ref{iwpdep}).

It may be expected that higher-dimensional generalizations of these
NAF black holes to NAF static black brane solutions to the
Einstein-$p$-form-dilaton theory also exist, as well as NAF rotating and
dyonic black branes for special values of the dilaton coupling
constant. We intend to address this question in a future
publication \cite{nbb}.

\section*{Appendix A}
\def\theequation{A.\arabic{equation}}
\setcounter{equation}{0}
Here we derive from the trivially embedded Taub-NUT solution a NUTted
NAF black hole of the electric sector, and show that its magnetic dual
is a twisted dimensional reduction of the Myers-Perry black hole with
two equal angular momenta. The trivial five-dimensional embedding of
the Taub-NUT metric is
\be
ds_5^2 = -\frac{\Delta}{\Sigma}\bigg(dt' +
2l\cos\theta\,d\varphi\bigg)^2 \nonumber 
+ \frac{\Sigma}{\Delta}\bigg(dr^2 + \Delta\,d\Omega_2^2\bigg) +
(dx^{'5})^2\,,
\ee 
with 
\be
\Delta = r^2 - 2mr - l^2\,, \quad \Sigma = r^2 + l^2\,.
\ee
The twist
\be 
t' = \gamma^{-1}(t-\gamma^2x^5)\,, \quad x^{'5} = \gamma x^5\,,
\ee
leads to the five-dimensional metric
\ba
ds_5^2 & = & -\frac{\Delta}{\Pi}\bigg(dt +
2\gamma l\cos\theta\,d\varphi\bigg)^2  
+ \frac{\Sigma}{\Delta}\bigg(dr^2 + \Delta\,d\Omega_2^2\bigg)
\nonumber \\ && + \frac{\Pi}{\Sigma}\bigg(dx^5 + \frac{\Delta}{\Pi}\bigg[dt +
2\gamma l\cos\theta\,d\varphi\bigg]\bigg)^2\,,
\ea 
with
\be
\Pi = 2\gamma^2(mr + l^2)\,.
\ee
After dimensional reduction, this leads to the four-dimensional NAF
``electric'' solution
\ba
ds_4^2 & = & - \frac{\Delta}{\sqrt{\Pi\Sigma}}\bigg(dt + 2\gamma l\cos\theta
\,d\varphi\bigg)^2 +\sqrt{\Pi\Sigma} \bigg(\frac{dr^2}{\Delta} +
d\Omega_2^2 
\bigg)\,, \lb{nerotg} \\ \A & = &
\frac{\Delta}{2\Pi}\bigg(dt + 2\gamma l\cos\theta\,d\varphi\bigg), \lb{nerota}
\\ \e^{2\phi/\sqrt3} & = & \sqrt{\frac{\Sigma}{\Pi}}\,,\lb{nerotp}
\ea

The four-dimensional ``magnetic'' dual solution has the same
four-dimensional metric with electromagnetic and dilaton fields
\ba
\A & = & -\frac{\gamma}{\Sigma}\bigg(l(r-m)\,dt + \frac{\gamma
  m\Lambda}{\Gamma} \cos\theta\,d\varphi\bigg), \lb{nmrota}
\\ \e^{2\phi/\sqrt3} & = & \sqrt{\frac{\Pi}{\Sigma}}
\qquad (\Lambda = r^2 + 2l^2r/m -l^2)\,.\lb{nmrotp}
\ea
The corresponding five-dimensional lift
\ba
ds_5^2 & = & -\frac{\Delta}{\Gamma}\bigg(dt +
2\gamma l\cos\theta\,d\varphi\bigg)^2  
+ \frac{\Pi}{\Delta}\bigg(dr^2 + \Delta\,d\Omega_2^2\bigg)
\nonumber \\ && + \frac{\Sigma}{\Pi}\bigg(dx^5 
- 2\frac{\gamma}{\Sigma}\bigg[l(r-m)dt + \frac{\gamma
  m\Lambda}{\Gamma} \cos\theta\,d\varphi\bigg]\bigg)^2\,,
\ea 
may be rearranged as
\ba
&& ds_5^2 = -\bigg(dt + \bar{a}\,d\eta\bigg)^2 +
\frac{\mu}{\rho^2}\bigg(dt + \bar{a}\,d\eta - \frac{\bar{a}}2(d\eta -
  \cos\theta\,d\varphi)\bigg)^2 \nonumber \\
& & + \frac{\rho^4}{\rho^4 - \mu\rho^2 + \mu\bar{a}^2}\,d\rho^2 +
  \frac{\rho^2}4\bigg(d\theta^2 + d\varphi^2 + d\eta^2 -
2\cos\theta\,d\varphi\,d\eta\bigg) \,, \lb{tmp}
\ea
with 
\be
\rho^2 = 8\gamma^2(mr + l^2)\,, \; \eta = \frac1{2m\gamma^2}\,x^5\,,
\; \mu = 16\gamma^2(l^2 + m^2)\,, \; \bar{a} = 2\gamma l\,.  
\ee
After carrying out the inverse twist
\be
t = t' - \bar{a}\eta\,,
\ee
we recognize in (\ref{tmp}) another form of the Myers-Perry metric with
two equal angular momenta, Eq. (\ref{mp}) with $a_+ = a_- = \bar{a}$.

\newpage

\begin{figure}
\centerline{\epsfxsize=400pt\epsfbox{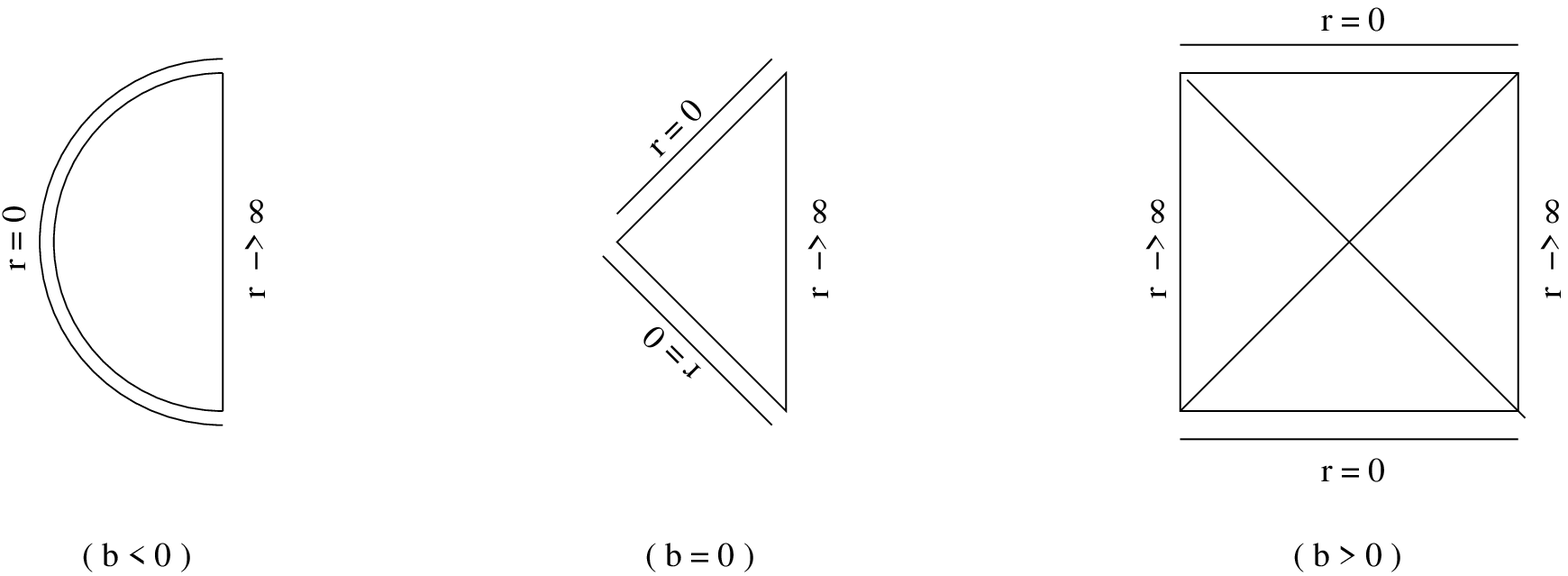}} \vspace{2cm}
\caption{Penrose diagrams of (\ref{nbhg}) for 
$0<\alpha<1$ with $b<0$, $b=0$ and $b>0$.}
\end{figure}

\begin{figure}
\centerline{\epsfxsize=400pt\epsfbox{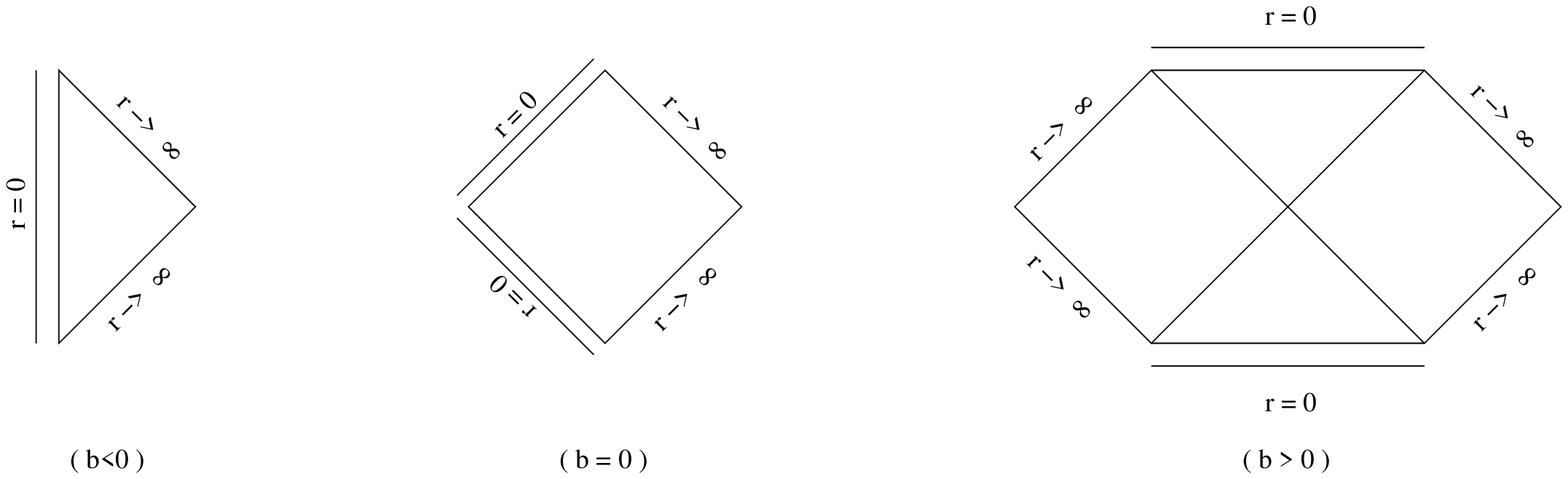}} \vspace{2cm}
\caption{Penrose diagrams of (\ref{nbhg}) for 
$\alpha=1$ with $b<0$, $b=0$ and $b>0$.}
\end{figure}

\begin{figure}
\centerline{\epsfxsize=400pt\epsfbox{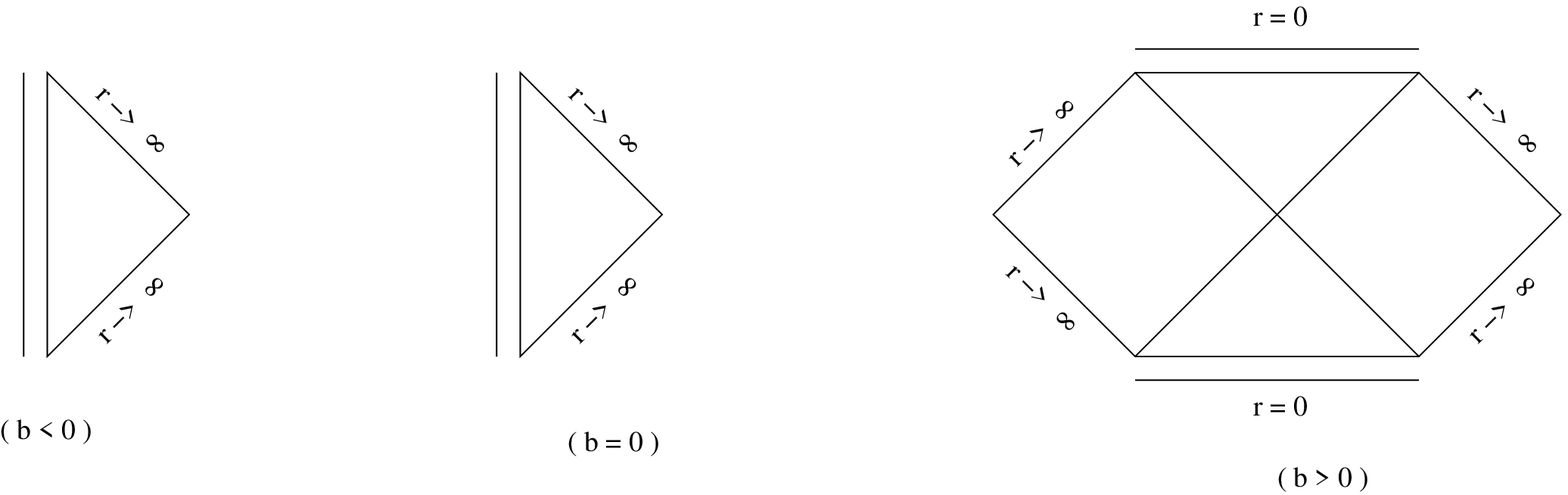}} \vspace{2cm}
\caption{Penrose diagrams of (\ref{nbhg}) for 
$\alpha>1$ with $b<0$, $b=0$ and $b>0$.}
\end{figure}

\end{document}